\def\be{\begin{equation}}
\def\ee{\end{equation}}
\def\ba{\begin{array}}
\def\ea{\end{array}}

\documentclass[aps,amsmath,amssymb,amsfonts,showpacs,pra]{revtex4}
\usepackage{graphicx,color}
\usepackage{amsmath,amsthm}
\def\qed{\leavevmode\unskip\penalty9999 \hbox{}\nobreak\hfill
     \quad\hbox{\leavevmode  \hbox to.77778em{%
               \hfil\vrule   \vbox to.675em%
               {\hrule width.6em\vfil\hrule}\vrule\hfil}}
     \par\vskip3pt}

\begin{document}
\title{Detecting high-dimensional multipartite entanglement via some classes of measurements}
\author{Lu Liu}
\author{Ting Gao}
\email{gaoting@hebtu.edu.cn}
\affiliation{College of Mathematics and Information Science, Hebei
Normal University, Shijiazhuang 050024, China}

\author{Fengli Yan}
\email{flyan@hebtu.edu.cn}
\affiliation{College of Physics Science and Information Engineering, Hebei
Normal University, Shijiazhuang 050024, China}

\begin{abstract}
Mutually unbiased bases, mutually unbiased measurements and general symmetric informationally complete  measurements are three related concepts in quantum information theory.
We investigate multipartite systems using these notions and present some criteria detecting entanglement of arbitrary high dimensional multi-qudit systems and multipartite systems of subsystems with different dimensions.
It is proved that these criteria can detect the $k$-nonseparability ($k$ is even) of multipartite qudit systems and arbitrary high dimensional multipartite systems of $m$ subsystems with different dimensions. We show that they are more efficient and wider of application range than the previous ones.
They provide experimental implementation in detecting entanglement without full quantum state tomography.
\end{abstract}


\maketitle

\section{Introduction}

Quantum entanglement is one of the most intriguing features of quantum mechanics which lies at the heart of quantum information sciences \cite{Horodecki09,Nielsen}. It has wide applications in diverse fields ranging from condensed matter \cite{RMP80.517} to high-energy field theory \cite{JPA42.504005}. The separability problem, namely distinguishing separable states from entangled states, is a challenging task whose complexity scales very unfavorably with the size of the system \cite{Guhne09}. A possible approach is to consider sufficient criteria for entanglement.
For bipartite systems, various separability criteria have been proposed.

When it comes to multipartite systems, the situation becomes much more complicated, because there exhibits much richer structure than bipartite case \cite{Guhne09}. Although the detection of multipartite entangled states is a harder challenge, it is worthy of study because they have advantages when performing some tasks \cite{Nature464.1165}. Many attempts have been made to frame multipartite entanglement detection, such as witnessing genuine multipartite entanglement \cite{PRL.113.100501,PRA.82.062113,NJP.12.053002}, detecting $k$-nonseparable states \cite{QIC.10.829,PRL.112.180501,EPL.104.20007,PRA.93.042310}, etc.
The main challenge for high-dimensional multipartite systems is not only to develop mathematical tools for entanglement detection, but also to find schemes whose experimental implementation requires minimal effort, that is to say, we need to detect entanglement with as few measurements as possible, specifically independent of full state tomography.

The notion of mutually unbiased bases (MUBs) was first introduced under a different name \cite{1409.3386[7]}.
Many quantum information protocols depend upon the use of MUBs \cite{PRL.110.143601}, such as quantum key distribution, the reconstruction of quantum states, etc.
The concept of MUBs was generalized to mutually unbiased measurements (MUMs) \cite{NJP.16.053038} due to the open problem of the maximum number of MUBs for non-prime power dimensions which limits its applications \cite{1409.3386[11]}. The construction of a complete set of $d+1$ MUMs were found \cite{NJP.16.053038} in a finite, $d$-dimensional Hilbert space, no matter whether $d$ is a prime power.
Symmetric informationally complete positive operator-valued measures (SIC-POVMs) is another related topic in quantum information, which has many helpful connections with MUBs, such as operational link \cite{PRA.88.032312}, quantum state tomography \cite{PRL.105.030406,PRA.83.052332,JMP.45.2171} and uncertainty relations \cite{EPJ.67.269}. Since it has some similar limitation as MUBs, in Ref.\cite{1305.6545}, the authors generalized the concept of SIC-POVMs to general symmetric informationally complete measurements (GSIC-POVMs), which were constructed without requiring to be rank one.

These quantum measurements have been used to detect entanglement recently. In Ref.\cite{PRA.86.022311}, the authors availed of MUBs and obtained separability criteria in arbitrarily high-dimensional quantum systems. Later some separability criteria for $d$-dimensional bipartite systems using MUMs were constructed \cite{PRA.89.064302,1407.7333}.
We obtained separability criteria for high dimensional and multipartite systems via MUMs \cite{SR}. A separability criterion for $d$-dimensional bipartite systems using GSIC-POVMs was given in Ref.\cite{1406.7820}.
Recently, Shen, Li and Duan proposed three separability criteria for $d$-dimensional bipartite quantum systems via the MUBs, MUMs and GSIC-POVMs, which are said more powerful than the corresponding ones above \cite{PRA.91.012326}.

The separability criteria mentioned above are of practical significance as they provide experimental implementation in detecting entanglement of unknown quantum states which only depend on some local measurements. What's more, they are efficient of characterizing bipartite entangled states. But few of them are referred to entanglement for multipartite systems, which are significant both in theoretical and experimental aspects.

In this paper, we study the separability problem and propose some criteria detecting entanglement for multipartite qudit systems and multipartite systems of different dimensional multi-level subsystems via MUBs, MUMs, and GSIC-POVMs. We show that the criteria in Ref.\cite{PRA.91.012326} are the special case of ours for two-qudit systems. What's more, our criteria can be applied to a wider range of multipartite systems and are more efficient than the former ones for multipartite systems. They also provide experimental implementation using only some local measurements.

\section{PRELIMINARIES}

The details of the notions of MUBs, MUMs, and GSIC-POVMs have been studied in original works \cite{1409.3386[11],NJP.16.053038,1305.6545}. Here we will briefly review the definitions of these measurements and the notions of $k$-separability and an operator used in the sequel.

Two orthonormal bases in Hilbert space $\mathbb{C}^{d}$ are called \emph{mutually unbiased} if and only if the transition probability from any state of one basis to any state of the second basis is constant. 
 A set of orthonormal bases $\{\mathcal{B}_{1},\mathcal{B}_{2},\cdots,\mathcal{B}_{m}\}$ of Hilbert space $\mathbb{C}^{d}$
 is called a set of \emph{mutually unbiased bases} (MUBs) if and only if every pair of bases in the set is mutually unbiased. The upper bound $d+1$ of the number of MUBs can be reached when $d$ is a prime power, but for even $d=6$ the maximal number of MUBs is an open problem \cite{1409.3386[11]}.

Kalev and Gour generalized the concept of MUBs to mutually unbiased measurements \cite{NJP.16.053038} which include the complete set of MUBs as a special case.
Two measurements on a $d$-dimensional Hilbert space,
${\mathcal{P}}^{(b)}=\{P^{(b)}_n|P^{(b)}_n\geq0,~\sum_{n=1}^{d}P^{(b)}_n=I\}$, $b{=}1,~2$, with $d$ elements each, are said to be \emph{mutually unbiased measurements} (MUMs) \cite{NJP.16.053038} if and only if
\begin{equation}
\begin{split}
\mathrm{Tr}(P^{(b)}_n)=&1,\\
\mathrm{Tr}(P^{(b)}_n P^{(b')}_{n'})=&\delta_{n,n'}\delta_{b,b'}\kappa
+(1-\delta_{n,n'})\delta_{b,b'}\frac{1-\kappa}{d-1}
+(1-\delta_{b,b'})\frac1{d},\\
\end{split}
\end{equation}
where $\kappa$ is efficiency parameter ($\frac{1}{d}<\kappa\leq 1$), and $\kappa=1$ if and only if all  $P^{(b)}_n$ are rank one projectors, i.e., ${\mathcal{P}}^{(1)}$ and ${\mathcal{P}}^{(2)}$ are given by MUBs.

A set $\{P_j| P_j\geq 0,\sum_{j=1}^{d^2}P_j=I_d\}$ with $d^{2}$ rank one operators acting on $\mathbb{C}^{d}$ is called to be symmetric informationally complete (SIC) positive operator valued measurements (POVMs), if the operator $\{P_j\}$ are of the form
$
P_{j}=\frac{1}{d}|\phi_{j}\rangle\langle\phi_{j}| ~ (j=1,2,\ldots,d^{2}),
$
where the vectors $|\phi_{j}\rangle$ satisfies
$
|\langle\phi_{j}|\phi_{k}\rangle|^{2}=\frac{1}{d+1},  j\neq k
$
\cite{1406.7820}.
Whether there exist SIC-POVMs in arbitrary dimension is still unknown \cite{1406.7820}.
In Ref.\cite{1305.6545}, the notion of SIC-POVMs was generalize to general symmetric informationally complete (GSIC) positive operator valued measurements (POVMs). A set of $d^{2}$ positive-semidefinite operators $\{P_{\alpha}\}_{\alpha=1}^{d^{2}}$ is a GSIC-POVM if and only if
\begin{equation}
\begin{split}
\sum_{\alpha=1}^{d^{2}}P_{\alpha}&=I,\\
\mathrm{Tr}(P_{\alpha}^{2})&=a,\\
\mathrm{Tr}(P_{\alpha}P_{\beta})&=\frac{1-da}{d(d^{2}-1)}, \forall\alpha,\beta\in\{1,2,\ldots,d^{2}\},\alpha\neq\beta,\\
\end{split}
\end{equation}
where $I$ denotes the identity operator, the parameter $a$ satisfies $\frac{1}{d^{3}}<a\leq\frac{1}{d^{2}}$. When the parameter satisfies $a=\frac{1}{d^{2}}$, all the operators $P_{\alpha}$ are rank one, that means $\{P_{\alpha}\}$ are given by SIC-POVMs \cite{1305.6545}.

For multipartite systems, there are various kinds of classification for multipartite entanglement. Next we introduce the notion of $k$-separable state for later use.
An $N$-partite system is described by Hilbert space $\mathcal{H}= \mathcal{H}_1\otimes \mathcal{H}_2\otimes\cdots\otimes\mathcal{H}_N$, where the dimension of the subspace $\mathcal{H}_i$ is denoted by $d_i$. A collection of pairwise disjoint sets $A_{1},A_{2},\cdots,A_{k}$ satisfying $\bigcup_{i=1}^{k}A_{i}=\{1,2,\ldots,N\}$ is defined as a $k$-partition $A_{1}|A_{2}|\cdots|A_{k}$.
A pure state $|\varphi\rangle$ of an $N$-partite quantum system is called $k$-separable if there exists a $k$-partition $A_{1}|A_{2}|\cdots|A_{k}$ such that the state can be written as a tensor product $|\varphi\rangle=|\varphi\rangle_{A_{1}}|\varphi\rangle_{A_{2}}\cdots|\varphi\rangle_{A_{k}}$, where $|\varphi\rangle_{A_{i}}$ is the state of subsystem $A_{i}$. A general mixed state $\rho$ is $k$-separable if it can be written as a mixture of $k$-separable states $\rho=\sum_{i}p_{i}\rho_{i}$, where $\rho_{i}$ is $k$-separable pure states \cite{EPL.104.20007}. States that are not $k$-separable are called $k$-nonseparable. In particular, $N$-separable states are fully separable, and states which are not bi-separable are genuinely $N$-partite entangled. In this paper, we consider $k$-separable mixed states as a convex combination of $N$-partite pure states, each of which is $k$-separable with respect to a fixed partition.

When $N$ is an even number, there are two different classes of
bipartite partitions $\mathcal{P_{I}}$ and $\mathcal{P_{II}}$
introduced in Ref.\cite{PRA.74.042338}. $\mathcal{P_{I}}$ denotes that both sides
of bipartite partition contain odd number of parties, and
$\mathcal{P_{II}}$ means even-number parties in each side.
For instance,
$\mathcal{P_{I}}=\{\rho_{1}\otimes\rho_{234},\rho_{2}\otimes\rho_{134},
\rho_{3}\otimes\rho_{124},\rho_{4}\otimes\rho_{123}\}$ and
$\mathcal{P_{II}}=
\{\rho,\rho_{12}\otimes\rho_{34},\rho_{13}\otimes\rho_{24},\rho_{14}\otimes\rho_{23}\}$
when $N=4$ \cite{PRA.77.060301}. An operator of their linear combination can
be defined \cite{PRA.77.060301}
\begin{equation}\label{deltarou}
    \Delta\rho=\frac{1}{2^{N-2}}(\mathcal{Q_{II}}-\mathcal{Q_{I}}),
\end{equation}
where $\mathcal{Q_{II}}=\sum_{q\in\mathcal{P_{II}}}q$ and
$\mathcal{Q_{I}}=\sum_{p\in\mathcal{P_{I}}}p$. For $N=2$ and $4$,
$\Delta\rho=\rho-\rho_{1}\otimes\rho_{2}$ and
$\frac{1}{4}(\rho+\rho_{12}\otimes\rho_{34}+\rho_{13}\otimes\rho_{24}+\rho_{14}\otimes\rho_{23}-
\rho_{1}\otimes\rho_{234}-\rho_{2}\otimes\rho_{134}-\rho_{3}\otimes\rho_{124}-\rho_{4}\otimes\rho_{123})$,
respectively. In the following, we will present separability criteria based on $\Delta\rho$.

\section{DETECTION OF MULTIPARTITE ENTANGLEMENT}

In this section, we present three separability criteria using  MUBs, MUMs, and GSIC-POVMs. Inspired by the operator (\ref{deltarou}) defined in Ref.\cite{PRA.77.060301}, in order to detect $k$-nonseparable states on multipartite systems of subsystems with different dimensions, we obtain the separability criterion in Hilbert space $\mathbb{C}^{d_{1}}\otimes\mathbb{C}^{d_{2}}\otimes\cdots\otimes\mathbb{C}^{d_{m}} $.

\vspace{0.2cm}{\slshape Theorem 1.} Let $\{\mathcal{B}_{j,1},\mathcal{B}_{j,2},\cdots,\mathcal{B}_{j,M_j}\}$ be a set of MUBs in $\mathbb{C}^{d_{j}}$, where $j=1,2,\cdots,m$. Define
\begin{align}
L(\rho)&=\max_{\begin{subarray}{c} \{|i_{j,k}\rangle\}\subseteq \mathcal{B}_{j,k} \end{subarray}}
         \sum\limits_{k=1}^{M}\sum\limits_{i=1}^{d}
         |\langle i_{1,k}i_{2,k}\cdots i_{m,k}|\Delta\rho|i_{1,k}i_{2,k}\cdots i_{m,k}\rangle|,
\end{align}
where  $d=\min\{d_{1},d_{2},\cdots,d_{m}\}$, $M=\min\{M_{1},M_{2},\cdots,M_{m}\}$, $\Delta\rho$ is the operator defined as (\ref{deltarou}), and $m$ is an even number.   If a state $\rho$ in $\mathbb{C}^{d_{1}}\otimes\mathbb{C}^{d_{2}}\otimes\cdots\otimes\mathbb{C}^{d_{m}} $ is fully separable, then
\begin{align}
L(\rho)\leq\min_{1\leq a\neq b\leq m}
         \sqrt{1+\frac{M_{a}-1}{d_{a}}-\sum\limits_{k=1}^{M_a}\sum\limits_{i=1}^{d_a}\langle i_{a,k}|\rho^{a}|i_{a,k}\rangle^{2}}
         \sqrt{1+\frac{M_{b}-1}{d_{b}}-\sum\limits_{k=1}^{M_b}\sum\limits_{i=1}^{d_b}\langle i_{b,k}|\rho^{b}|i_{b,k}\rangle^{2}},
\end{align}
where $\rho^a$ ($\rho^b$) is the reduced density matrix of the $a$-th ($b$-th) subsystem.
\vspace{0.2cm}

{\slshape Proof.}~Any fully separable state $\rho$ can be written as $\rho=\sum_{i}p_{i}\rho_{i}
^{1}\otimes\rho_{i}^{2}\otimes\cdots\otimes\rho_{i}^{m}$, where
$\{p_{i}\}$ is a probability distribution and
 $\rho_{i}^{k}$ denotes the pure state density matrix acting on the $k$-th subsystem. By
\begin{equation}\label{X}
\begin{array}{ll}
\Delta\rho & =\frac{1}{2^{m-2}}(\mathcal{Q_{II}}-\mathcal{Q_{I}})\\
&=\frac{1}{2^{m-1}}\sum_{k,l}p_{k}p_{l}(\rho_{k}^{1}-\rho_{l}^{1})
\otimes(\rho_{k}^{2}-\rho_{l}^{2})\otimes\cdots\otimes(\rho_{k}^{m}-\rho_{l}^{m}),
\end{array}
\end{equation}
given in Ref.\cite{PRA.77.060301}, we have

\begin{align*}
     & \sum\limits_{k=1}^{M}\sum\limits_{i=1}^{d}
         |\langle i_{1,k}i_{2,k}\cdots i_{m,k}|\Delta\rho|i_{1,k}i_{2,k}\cdots i_{m,k}\rangle|\\
\leq & 2\sum\limits_{k=1}^{M}\sum\limits_{i=1}^{d}
        \sum_{r,s}p_{r}p_{s}\prod_{t=1}^{m}\frac{|\langle i_{t,k}|(\rho_{r}^{t}-\rho_{s}^{t})|i_{t,k}\rangle|}{2},
   \end{align*}
and $0\leq |\frac{\langle i_{t,k}|(\rho_{r}^{t}-\rho_{s}^{t})|i_{t,k}\rangle}{2}|\leq 1$. For arbitrary $1\leq a\neq b\leq m$, we get
\begin{align*}
& \sum\limits_{k=1}^{M}\sum\limits_{i=1}^{d}
         |\langle i_{1,k}i_{2,k}\cdots i_{m,k}|\Delta\rho|i_{1,k}i_{2,k}\cdots i_{m,k}\rangle| \\
 \leq &2\sum\limits_{k=1}^{M}\sum\limits_{i=1}^{d}\sum\limits_{r,s}
              \sqrt{p_{r}p_{s}}\frac{|\langle i_{a,k}|(\rho_{r}^{a}-\rho_{s}^{a})|i_{a,k}\rangle|}{2}
              \sqrt{p_{r}p_{s}}\frac{|\langle i_{b,k}|(\rho_{r}^{b}-\rho_{s}^{b})|i_{b,k}\rangle|}{2}\\
\leq & 2 \sqrt{\sum\limits_{k=1}^{M}\sum\limits_{i=1}^{d}\sum\limits_{r,s}p_{r}p_{s}
                   \Big[\frac{\langle i_{a,k}|(\rho_{r}^{a}-\rho_{s}^{a})|i_{a,k}\rangle}{2}\Big]^2}
              \sqrt{\sum\limits_{k=1}^{M}\sum\limits_{i=1}^{d}\sum\limits_{r,s}p_{r}p_{s}
                   \Big[\frac{\langle i_{b,k}|(\rho_{r}^{b}-\rho_{s}^{b})|i_{b,k}\rangle}{2}\Big]^2}\\
   = & \sqrt{\sum\limits_{k=1}^{M}\sum\limits_{i=1}^{d}
              [\sum\limits_{r}p_{r}\langle i_{a,k}|\rho_{r}^{a}|i_{a,k}\rangle^{2}-\langle i_{a,k}|\rho^{a}|i_{a,k}\rangle^{2}]}
        \sqrt{\sum\limits_{k=1}^{M}\sum\limits_{i=1}^{d}
              [\sum\limits_{r}p_{r}\langle i_{b,k}|\rho_{r}^{b}|i_{b,k}\rangle^{2}-\langle i_{b,k}|\rho^{b}|i_{b,k}\rangle^{2}]}\\
\leq & \sqrt{1+\frac{M_{a}-1}{d_{a}}-\sum\limits_{k=1}^{M}\sum\limits_{i=1}^{d}
           \langle i_{a,k}|\rho^{a}|i_{a,k}\rangle^{2}}
      \sqrt{1+\frac{M_{b}-1}{d_{b}}-\sum\limits_{k=1}^{M}\sum\limits_{i=1}^{d}
           \langle i_{b,k}|\rho^{b}|i_{b,k}\rangle^{2}}.,
\end{align*}
where Cauchy-Schwarz inequality is used in the second inequality, and the last inequality is due to the relation \cite{PRA.79.022104}
\begin{equation}
\sum\limits_{k=1}^{M}\sum\limits_{i=1}^{d}\langle i_{k}|\rho|i_{k}\rangle^{2}\leq1+\frac{M-1}{d},
\end{equation}
for any pure state $\rho$ in $\mathbb{C}^{d}$.
Because of the arbitrariness of $a,b$, we complete the proof.
\hfill $\square$

 It is worthy to note that Theorem 1 in Ref. \cite{PRA.91.012326} is the corollary of Theorem 1. When $m$=2, $d_1=d_2=d$, and $M_1=M_2=M$, this Theorem 1 reduced to Theorem 1 in Ref. \cite{PRA.91.012326}. It was shown that Theorem 1 in Ref. \cite{PRA.91.012326} is stronger than the MUB criterion in Ref. \cite{PRA.86.022311}. As the special case of our criterion, Theorem 1 in Ref. \cite{PRA.91.012326} can only be applied to bipartite systems of two $d$-dimensional subsystems, it is clear that our criterion of Theorem 1 is more effective than the MUB criterion in Ref. \cite{PRA.86.022311} and wider application range than Theorem 1 in Ref. \cite{PRA.91.012326}.

Next, we present the separability criteria using MUMs and GSIC-POVMs, which are more powerful than that via MUBs due to the fact that the complete set of MUMs and GSIC-POVMs always exist in all finite dimensions.

\vspace{0.2cm}{\slshape Theorem 2.}
Suppose that $\rho$ is a density matrix in $\mathbb{C}^{d_{1}}\otimes\mathbb{C}^{d_{2}}\otimes\cdots\otimes\mathbb{C}^{d_{m}} $ and $\mathcal{P}^{(b)}_{i}$ are any sets of $M$ MUMs on $\mathbb{C}^{d_{i}}$ with the efficiencies $\kappa_{i}$,
where  $b=1,2,\cdots,M$, $i=1,2,\cdots,m$.
Let $d=\min \{d_{1},d_{2},\cdots,d_{m}\},$ and
$$S(\rho)=\max_{\begin{subarray}{c} \{P_{i,n}^{(b)}\}_{n=1}^{d}\subseteq\mathcal{P}^{(b)}_{i} \\ i=1,2,\cdots,m \\ b=1,2,\cdots,M \end{subarray}}
\sum\limits_{b=1}^{M}\sum\limits_{n=1}^{d}
\big|\textrm{Tr}\Big(\big(\otimes^{m}_{i=1} P_{i,n}^{(b)}\big) \Delta\rho\Big)\big|.$$
For even number $m$, if $\rho$ is fully separable, then
\begin{align} \label{M-2}
S(\rho)\leq\min_{1\leq i\neq j\leq m}\sqrt{(\frac{M-1}{d_{i}}+\kappa_{i})-\sum\limits_{b=1}^{M}\sum\limits_{n=1}^{d}
              [\textrm{Tr}(P_{i,n}^{(b)}\rho^{i}\big)]^{2}}
              \sqrt{(\frac{M-1}{d_{j}}+\kappa_{j})-\sum\limits_{b=1}^{M}\sum\limits_{n=1}^{d}
              [\textrm{Tr}(P_{j,n}^{(b)}\rho^{j}\big)]^{2}}.
\end{align}
Here $\rho^i$ ($\rho^j$) is the reduced density matrix of the $i$-th ($j$-th) subsystem.

\vspace{0.2cm}
{\slshape Proof.}~
Since $\Delta\rho$ can be written in the form (\ref{X}), for arbitrary $1\leq i\neq j\leq m$, we obtain
\begin{equation*}
\begin{array}{ll}
&\sum\limits_{b=1}^{M}\sum\limits_{n=1}^{d}
              \big|\textrm{Tr}[(\otimes^{m}_{i=1} P_{i,n}^{(b)})\Delta\rho]\big|\\
\leq&\sum\limits_{b=1}^{M}\sum\limits_{n=1}^{d}
              \sum\limits_{kl}2p_{k}p_{l}\prod_{i=1}^{m}\big|\frac{1}{2}\textrm{Tr}[P_{i,n}^{(b)}(\rho^{i}_{k}-\rho^{i}_{l})]\big|\\
\leq&\frac{1}{2}\sum\limits_{b=1}^{M}\sum\limits_{n=1}^{d}\sum\limits_{kl}p_{k}p_{l}
                        |\textrm{Tr}(P_{i,n}^{(b)}(\rho^{i}_{k}-\rho^{i}_{l}))|
                        |\textrm{Tr}(P_{j,n}^{(b)}(\rho^{j}_{k}-\rho^{j}_{l}))|\\
\leq&\frac{1}{2}\sqrt{\sum\limits_{b=1}^{M}\sum\limits_{n=1}^{d}\sum\limits_{kl}p_{k}p_{l}
                        [\textrm{Tr}(P_{i,n}^{(b)}(\rho^{i}_{k}-\rho^{i}_{l}))]^{2}}
                \sqrt{\sum\limits_{b=1}^{M}\sum\limits_{n=1}^{d}\sum\limits_{kl}p_{k}p_{l}
                        [\textrm{Tr}(P_{j,n}^{(b)}(\rho^{j}_{k}-\rho^{j}_{l}))]^{2}}\\
=&\sqrt{\sum\limits_{b=1}^{M}\sum\limits_{n=1}^{d}\{\sum\limits_{k}p_{k}
                        [\textrm{Tr}(P_{i,n}^{(b)}\rho^{i}_{k}\big)]^{2}-[\textrm{Tr}(P_{i,n}^{(b)}\rho^{i}\big)]^{2}\}}
  \sqrt{\sum\limits_{b=1}^{M}\sum\limits_{n=1}^{d}\{\sum\limits_{k}p_{k}
                        [\textrm{Tr}(P_{j,n}^{(b)}\rho^{j}_{k}\big)]^{2}-[\textrm{Tr}(P_{j,n}^{(b)}\rho^{j}\big)]^{2}\}}\\
\leq&\sqrt{(\frac{M-1}{d_{i}}+\kappa_{i})-\sum\limits_{b=1}^{M}\sum\limits_{n=1}^{d}
              [\textrm{Tr}(P_{i,n}^{(b)}\rho^{i}\big)]^{2}}
              \sqrt{(\frac{M-1}{d_{j}}+\kappa_{j})-\sum\limits_{b=1}^{M}\sum\limits_{n=1}^{d}
              [\textrm{Tr}(P_{j,n}^{(b)}\rho^{j}\big)]^{2}},
\end{array}
\end{equation*}
where we have used Cauchy-Schwarz inequality, and inequality \cite{1407.7333}
\begin{equation}
\sum_{b=1}^M\sum_{n_i=1}^{d_i}[\textrm{Tr}(P_{i,n_i}^{(b)}\rho\big)]^{2}\leq\frac{M-1}{d_i}+\frac{1-\kappa_i+(\kappa_i d_i-1)\textrm{Tr}(\rho^{2})}{d_i-1},\label{M-1}
\end{equation}
for pure states $\rho^{i}_{k}$.
It is complete due to the the arbitrariness of $i,j$.
 \hfill$\square$

Note that when the conditions are limited to only two subsystems with the same dimension and the complete sets of MUMs, that is, $m=2$, $d_1=d_2=d$, $M=d+1$, the inequality (\ref{M-2}) becomes
$S(\rho)\leq \sqrt{(1+\kappa)-\sum\limits_{b=1}^{d+1}\sum\limits_{n=1}^{d}
              [\textrm{Tr}(P_{1,n}^{(b)}\rho^{1}\big)]^{2}}
              \sqrt{(1+\kappa)-\sum\limits_{b=1}^{d+1}\sum\limits_{n=1}^{d}
              [\textrm{Tr}(P_{2,n}^{(b)}\rho^{2}\big)]^{2}}$,
which was the criterion based on MUMs, Theorem 2 in  Ref.\cite{PRA.91.012326}. That is, Theorem 2 in  Ref.\cite{PRA.91.012326} is the special case of our Theorem 2. Since the criterion based on MUMs in Ref.\cite{PRA.91.012326} can only be used to $d$-dimensional bipartite systems and is stronger than the corresponding one in Ref.\cite{PRA.89.064302},  Theorem 2 is more effective than the MUM criterion in Ref.\cite{PRA.89.064302} and wider range of application than that in Ref.\cite{PRA.91.012326}.

\vspace{0.2cm} {\slshape Theorem 3.}
Let $\rho$ be a density matrix in $\mathbb{C}^{d_{1}}\otimes\mathbb{C}^{d_{2}}\otimes\cdots\otimes\mathbb{C}^{d_{m}} $ and $\mathcal{P}_{i}$ are any $m$ sets of general symmetric informationally complete measurements on $\mathbb{C}^{d_{i}}$ with the
parameters $a_{i}$,  where $i=1,2,\cdots,m$, and $m$ is even.
Define
$$R(\rho)=\max_{\begin{subarray}{c} \{P_{i,n}\}_{n=1}^{d^{2}}\subseteq\mathcal{P}_{i} \\ i=1,2,\cdots,m \end{subarray}}
\sum\limits_{n=1}^{d^{2}}\textrm{Tr}(\otimes^{m}_{i=1} P_{i,n}\Delta\rho),$$
where $d=\min \{d_{1},d_{2},\cdots,d_{m}\}.$
If $\rho$ is fully separable, then
\begin{align} \label{M-3}
R(\rho)\leq\min_{1\leq i\neq j\leq m}
     \sqrt{\frac{a_{i}d_{i}^{2}+1}{d_{i}(d_{i}+1)}
              -\sum\limits_{n=1}^{d^{2}}[\textrm{Tr}(P_{i,n}\rho^{i}\big)]^{2}}
     \sqrt{\frac{a_{j}d_{j}^{2}+1}{d_{j}(d_{j}+1)}
              -\sum\limits_{n=1}^{d^{2}}[\textrm{Tr}(P_{j,n}\rho^{j}\big)]^{2}}.
\end{align}
Here $\rho^i$ ($\rho^j$) is the reduced density matrix of the $i$-th ($j$-th) subsystem.

\vspace{0.2cm}
{\slshape Proof.}~
By an analogous argument as Theorem 3 we obtain
\begin{equation*}
\begin{array}{ll}
&\sum\limits_{n=1}^{d^{2}}\textrm{Tr}(\otimes^{m}_{i=1} P_{i,n}\Delta\rho)\\
=&\sum\limits_{n=1}^{d^{2}}\sum\limits_{kl}2p_{k}p_{l}\prod_{i=1}^{m}[\frac{1}{2}\textrm{Tr}(P_{i,n}(\rho^{i}_{k}-\rho^{i}_{l}))]\\
\leq&\sum\limits_{n=1}^{d^{2}}\sum\limits_{kl}2p_{k}p_{l}
              [\frac{1}{2}\textrm{Tr}(P_{i,n}(\rho^{i}_{k}-\rho^{i}_{l}))]
              [\frac{1}{2}\textrm{Tr}(P_{j,n}(\rho^{j}_{k}-\rho^{j}_{l}))]\\
\leq&\frac{1}{2}\sqrt{\sum\limits_{n=1}^{d^{2}}\sum\limits_{kl}p_{k}p_{l}
              [\textrm{Tr}(P_{i,n}(\rho^{i}_{k}-\rho^{i}_{l}))]^{2}}
              \sqrt{\sum\limits_{n=1}^{d^{2}}\sum\limits_{kl}p_{k}p_{l}
              [\textrm{Tr}(P_{j,n}(\rho^{j}_{k}-\rho^{j}_{l}))]^{2}}\\
=&\sqrt{\sum\limits_{n=1}^{d^{2}}\{\sum\limits_{k}p_{k}[\textrm{Tr}(P_{i,n}\rho^{i}_{k}\big)]^{2}
              -[\textrm{Tr}(P_{i,n}\rho^{i}\big)]^{2}\}}
              \sqrt{\sum\limits_{n=1}^{d^{2}}\{\sum\limits_{k}p_{k}[\textrm{Tr}(P_{j,n}\rho^{j}_{k}\big)]^{2}
              -[\textrm{Tr}(P_{j,n}\rho^{j}\big)]^{2}\}}\\
\leq&\sqrt{\frac{a_{i}d_{i}^{2}+1}{d_{i}(d_{i}+1)}
              -\sum\limits_{n=1}^{d^{2}}[\textrm{Tr}(P_{i,n}\rho^{i}\big)]^{2}}
     \sqrt{\frac{a_{j}d_{j}^{2}+1}{d_{j}(d_{j}+1)}
              -\sum\limits_{n=1}^{d^{2}}[\textrm{Tr}(P_{j,n}\rho^{j}\big)]^{2}},
\end{array}
\end{equation*}
where $1\leq i\neq j\leq m$, and we have used Cauchy-Schwarz inequality as well as inequality \cite{PS89.085101}
\begin{equation}
\sum\limits_{n_i=1}^{d_i^{2}}[\mathrm{Tr}(P_{i,n_i}\rho)]^{2}=\frac{(a_id^{3}-1)\mathrm{Tr}(\rho^{2})+d(1-a_id)}{d(d^{2}-1)} \label{SIC}
\end{equation}
for pure states $\rho^{i}_{k}$.  It is complete.   \hfill $\square$

When $m=2$ and $d_1=d_2=d$, Theorem 3 reduced to Theorem 2, the GSIC-POVMs criterion in Ref. \cite{PRA.91.012326}. As the GSIC-POVMs criterion in Ref. \cite{PRA.91.012326} is stronger than the one in Ref. \cite{1406.7820}, so is our Theorem 3. Furthermore, Theorem 3 is suitable for multipartite systems composed of subsystems with different dimensions, it is more effective than the corresponding one in Ref. \cite{1406.7820} and wider range of application than the separability criteria in Ref. \cite{PRA.91.012326}.

For Theorems 1, 2, and 3, the dimensions of subsystems are not required to be the same, so we can straightforwardly detect $k$-nonseparable states ($k$ is even) with respect to a fixed partition. The sets $S_{k}$ of all $k$-separable states with respect to a fixed partition have nested structure, that is, each set is embedded within the next set: $S_N\subset S_{N-1}\subset\cdots \subset S_2\subset S_1$, and the complement $S_1\setminus S_k$ of $S_k$ in $S_1$ is the set of all $k$-nonseparable states with respect to fixed partition. So if a multipartite state is $N$-nonseparable ($N$ is even) using our criteria, since we don't require each subsystem has the same dimension, we can go on detecting whether it is $(N-2)$-nonseparable and so on. In this way, we do not just detect a given state is entangled or not, we can obtained the ``degrees of entanglement" to some extent by the notion of $k$-nonseparability.

The criteria given by Theorems 1-3 are much better than the previous ones  in Ref.\cite{PRA.86.022311, PRA.89.064302, 1406.7820, PRA.91.012326}.  First, they are more powerful than the main result ( inequality (7) ) in Ref.\cite{PRA.86.022311}, the criterion in Ref.\cite{PRA.89.064302},  and the criterion in Ref.\cite{1406.7820}, respectively.
Criteria in Ref.\cite{PRA.91.012326} are the special cases of our criteria, thus, they are corollaries of ours. Since the criteria in Ref.\cite{PRA.91.012326}  are proved to be more powerful than the corresponding ones introduced previously in Ref.\cite{PRA.86.022311, PRA.89.064302, 1406.7820}, so is ours. Second, our criteria are of wider range of application than the corresponding ones in Ref.\cite{PRA.86.022311, PRA.89.064302, 1406.7820, PRA.91.012326}.
We present separability criteria to detect entanglement of quantum states in $\mathbb{C}^{d_{1}}\otimes\mathbb{C}^{d_{2}}\otimes\cdots\otimes\mathbb{C}^{d_{m}} $,
where  $m\geq 2$, while the corresponding ones in Ref.\cite{PRA.86.022311, PRA.89.064302, 1406.7820, PRA.91.012326}  are
only  for a bipartite system of two $d$-dimensional subsystems,  that is, the presented criteria can be used to not only bipartite systems of two subsystems with same dimension but also multipartite qudit systems  and multipartite systems of subsystems with different dimensions, while the corresponding criteria in Ref.\cite{PRA.86.022311, PRA.89.064302, 1406.7820, PRA.91.012326} are applied to bipartite systems of two subsystems with same dimension.

In recent years, people gradually recognized the significance of multipartite quantum states with higher dimensions. Maximally entangled qudits have been proved to violate local realism more strongly and are less affected by noise than qubits \cite{PRL.85.4418,PRL.88.040404,PRL.96.060406}. Qudit states also have benefit in quantum communication, since they are more secure against eavesdropping attacks and more reliable in quantum processing \cite{PRL.88.127902,PRA.75.022313}. From the experimental viewpoint, the entangled qudits can be physically realized in linear photon systems \cite{PRL.97.023602}, etc. So it is important to characterize entanglement in multipartite systems, and it is still under intensive research.

\section{Conclusion}
In this paper, MUBs, MUMs and GSIC-POVMs have been used to study the entanglement detection of arbitrary high dimensional multipartite systems. We have present separability criteria given in Theorems 1-3 to detect entanglement of quantum states in $\mathbb{C}^{d_{1}}\otimes\mathbb{C}^{d_{2}}\otimes\cdots\otimes\mathbb{C}^{d_{m}} $. The presented criteria have wider range of application than the corresponding ones in Ref.\cite{PRA.86.022311, PRA.89.064302, 1406.7820, PRA.91.012326}, and are more efficient than the main result ( inequality (7) ) in Ref.\cite{PRA.86.022311}, the criterion in Ref.\cite{PRA.89.064302},  and the criterion in Ref.\cite{1406.7820}, respectively. The criteria offered in \cite{PRA.91.012326} are the special cases of our criteria.
Compared with some other separability criteria, the criteria given in Theorems 1-3 provide experimental implementation in detecting entanglement of unknown quantum states, and one can flexibly use them in practice because they require only a few local measurements.
 We can detect the $k$-nonseparability ($k$ is even) of multipartite qudit systems and arbitrary high dimensional multipartite systems of $m$ subsystems with different dimensions.

\acknowledgments
This work was supported by the National Natural Science Foundation
of China under Grant Nos: 11371005, 11475054, Hebei Natural Science Foundation
of China under Grant No: A2016205145. \\

Author  contribution statement\\

L.L., T.G. and F.Y. contributed equally to this work.  All authors wrote the main manuscript text and reviewed the manuscript.

\end{document}